\input{aipcheck}

\documentclass[
    ,final            
  ]
  {aipproc}

\layoutstyle{6x9}

\begin{document}

\title{Formation and Evolution of Supermassive Black Holes in Galactic Centers: Observational Constraints}

\author{G\"unther Hasinger}{
address={Max-Planck-Institut f\"ur extraterrestrische Physik \\
Postfach 1319, D--84541 Garching, Germany}
}
\author{and the CDF-S team}{address={ }}

\begin{abstract}

Deep X--ray surveys have shown that the cosmic X--ray background (XRB) is largely due to the accretion onto supermassive black holes, integrated over the cosmic time. These surveys have resolved more than 80\% of the 0.1-10 keV X--ray background into discrete sources. Optical spectroscopic identifications show that the sources producing the bulk of the X--ray background are a mixture of obscured (type-1) and unobscured (type-2) AGNs, as predicted by the XRB population synthesis models. A class of highly luminous type-2 AGN, so called QSO-2s, has been detected in the deepest {\em Chandra} and {\em XMM-Newton} surveys. The new {\em Chandra} AGN redshift distribution peaks at much lower redshifts (z $\approx$ 0.7) than that based on ROSAT data, indicating that Seyfert galaxies peak at significantly lower redshifts than QSOs.  

\end{abstract}

\maketitle

\section{Introduction}

Deep X-ray surveys indicate that the cosmic X-ray background (XRB) is largely due to accretion onto supermassive black holes, integrated over cosmic time. In the soft (0.5-2 keV) band more than 90\% of the XRB flux has been resolved using 1.4 Msec observations with ROSAT \cite{has98} and recently 1-2 Msec Chandra observations \cite{ros02,bra02} and 100 ksec observations with XMM-Newton \cite{has01}. In the harder (2-10 keV) band a similar fraction of the background has been resolved with the above Chandra and XMM-Newton surveys, reaching source densities of about 4000 deg$^{-2}$. 

The X-ray observations have so far been about consistent with population synthesis models based on unified AGN schemes \cite{com95,gil01}, which explain the hard spectrum of the X-ray background by a mixture of absorbed and unabsorbed AGN, folded with the corresponding luminosity function and its cosmological evolution. According to these models, most AGN spectra are heavily absorbed and about 80\% of the light produced by accretion will be absorbed by gas and dust \cite{fab98}. In particular they require a substantial contribution of  high-luminosity obscured X-ray sources (type-2 QSOs), which so far have only scarcely been detected. 

Optical follow-up programs with 8-10m telescopes have been completed for the 
ROSAT deep surveys and find predominantly AGN counterparts of the faint X--ray source population \cite{schm98,leh01} mainly X--ray and optically unobscured AGN (type-1 Seyferts and QSOs) and a smaller fraction of obscured AGN (type-2 Seyferts). Optical identifications for the deepest {\em Chandra} and {\em XMM-Newton} fields are now approaching a completeness of 60-80\% and find a 
mixture of obscured and unobscured AGN with an increasing fraction of 
obscuration towards fainter fluxes \cite{bar02,szo03}. Interestingly, first examples of the long-sought class of type-2 QSO have been detected in deep {\em Chandra} fields \cite{nor02,ste02}.

After having understood the basic contributions to the X-ray background, the general interest is now focussing on understanding the physical nature of these sources, the cosmological evolution of their properties, and their role in models of galaxy evolution. We know that basically every galaxy with a spheroidal component in the local universe has a supermassive black hole in its centre \cite{geb00}. The luminosity function of luminous X-ray selected AGN shows strong cosmological density evolution at redshifts up to 2, which goes hand in hand with the evolution of optically selected QSO and radio quasars, as well as the cosmic star formation history \cite{miy00}. At the redshift peak of optically selected QSO around z=2 the AGN space density is several hundred times higher than locally, which is in line with the assumption that most  galaxies have been active in the past and that the feeding of their black holes is reflected in the X-ray background. While the comoving space density of optically and radio-selected QSO has been shown to decline significantly beyond a redshift of 2.7 \cite{ssg95,sha96,fan01}, the statistical quality of X-ray selected AGN high-redshift samples still needs to be improved \cite{miy00}. The new Chandra and XMM-Newton surveys are now providing strong additional constraints here.    

In this review we compare the optical identification work in the two deepest Chandra fields, the Chandra Deep Field South and the Hubble Deep Field North and show preliminary results on the cosmological evolution of Seyfert galaxies.

\section{The deepest Chandra Fields}

The Chandra X-ray Observatory has performed deep X-ray surveys in a number of fields with ever increasing exposure times \cite{mus00,hor00,gia01,ste02} and has  completed a 1 Msec exposure in the Chandra Deep Field South (CDF-S, \cite{ros02}) and a 2 Msec exposure in the Hubble Deep Field North (HDF-N, \cite{bra02}). In Figure 1 (left), we show the colour composite Chandra image of the CDF-S. This was constructed by combining images, smoothed with a Gaussian with $\sigma$=1'' in three bands (0.3-1 keV, 1-3 keV, 3-7 keV), which contain approximately equal numbers of photons from detected sources.  Blue sources are those undetected in the soft (0.5-2 keV) band, most likely due to intrinsic absorption from neutral hydrogen with column densities $N_H>10^{22}~cm^{-2}$. Very soft sources appear red. A few extended low surface brightness sources are also readily visible in the image. Figure 1 (right) shows a similar image for
the 970 ksec observation of the HDF-N, kindly supplied by N. Brandt \cite{bra01}. The corresponding 2 Msec image will be published soon.

The CDF-S was also observed with XMM-Newton for a net exposure of ~500 ksec in July 2001 and January 2002 (PI: J. Bergeron, see \cite{has02}). The EPIC cameras have a larger field-of-view than ACIS, and a number of new diffuse sources are detected just outside the Chandra image.  X-ray spectroscopy of a large number of sources will ultimately be very powerful with XMM-Newton (see \cite{mai02} for the Lockman Hole).

\begin{figure}
  \includegraphics[height=.3\textheight]{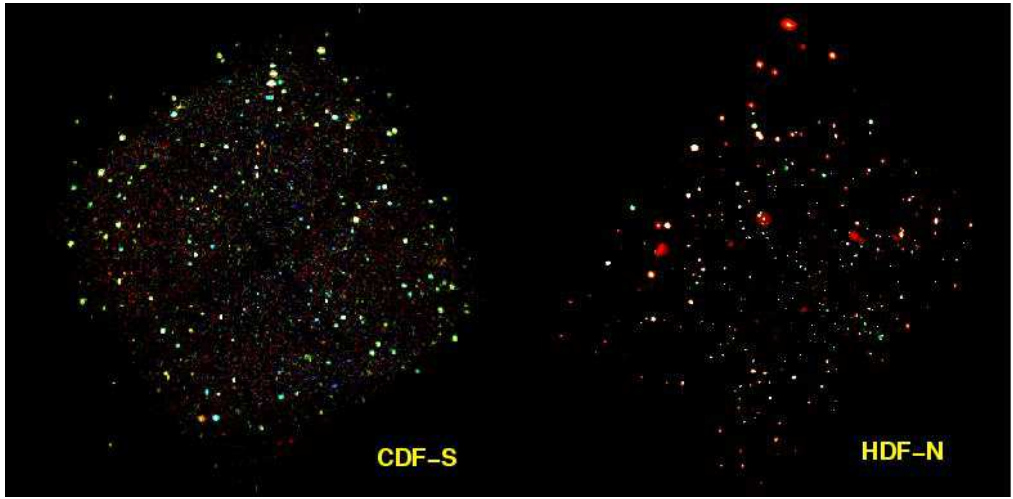}
  \caption{left: Color composite image of the Chandra Deep Field South of 940 ks (pixel size=0.984'', smoothed with a $\sigma$=1'' Gaussian).  The image was obtained combining three energy bands: 0.3-1 keV, 1-3 keV, 3-7 keV (respectively red, green and blue), from \cite{ros02}. Right: Similar image for the 970 ksec Chandra observation of the Hubble Deep Field North \cite{bra01}.}
\end{figure}

\section{VLT optical spectroscopy}

Optical spectroscopy in the CDF-S has been carried out in ~11 nights with the ESO Very Large Telescope (VLT) in the time frame April 2000 - December 2001, using deep optical imaging and low resolution multiobject spectroscopy with the FORS instruments with individual exposure times ranging from 1-5 hours. Some preliminary results including the VLT optical spectroscopy have already been presented \cite{nor02,ros02}. The complete optical spectroscopy will be published in \cite{szo03}. 

Redshifts could be obtained so far for 169 of the 346 sources in the CDF-S, of which 123 are very reliable (high quality spectra with 2 or more spectral features), while the remaining optical spectra contain only a single emission line, or are of lower S/N. For objects fainter than R=24 reliable redshifts can be obtained if the spectra contain strong emission lines. For the remaining optically faint objects we have to resort to photometric redshift techniques. Nevertheless, for a subsection of the sample at off-axis angles smaller than 8 arcmin we obtain a spectroscopic completeness of about 60\%. Including photometric redshifts \cite{wol01,mai03} this completeness increases to $\approx$ 80\% for the CDF-S.  

\begin{figure}
  \includegraphics[height=.4\textheight]{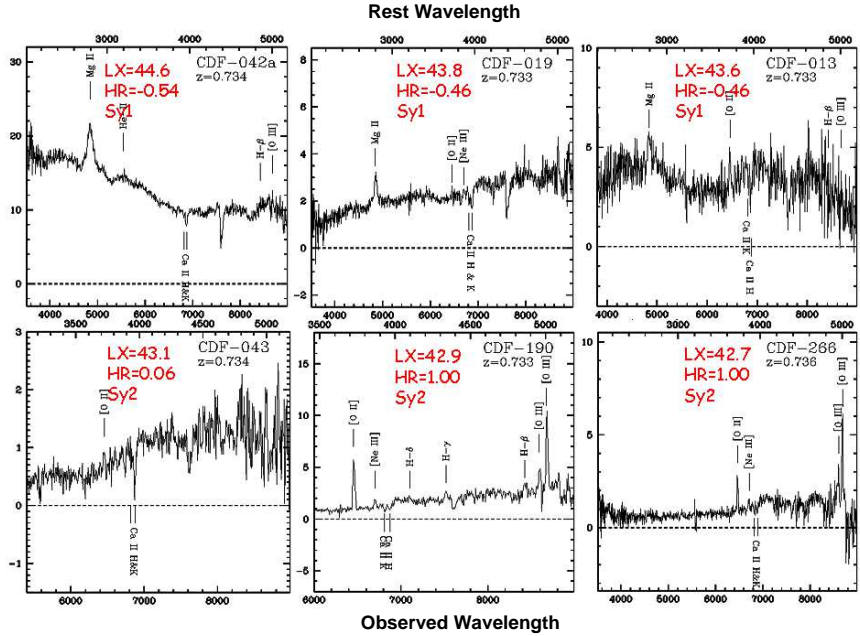}
  \caption{Optical spectra of six CDF-S sources selected from one of 
  the two observed redshift spikes at z=0.733 (see \cite{has02,gil03}), 
  obtained using multiobject-spectroscopy with FORS at the VLT \cite{szo03}. 
  The objects are sorted according to their X-ray luminosity, 
  which is given in each panel together with the X-ray hardness ratio and 
  the X-ray/optical classification.}
\end{figure}

Figure 2 shows examples of six VLT spectra of CDF-S sources of sources which are selected from one of the two redshift spikes detected in the field at z=0.733 \cite{has02,szo03,gil03}. The objects are sorted according to their X-ray luminosity which is given in the figure together with the X-ray hardness ratio ($HR=(H-S)/(H+S)$), where H and S are the count rates in the 2-7 keV and 0.5-2 keV, respectively. The spectrum in the upper left corresponds to a bona fide QSO, with a broad MgII line and a blue nonthermal continuum. This object has a high X-ray luminosity and soft X-ray spectrum consistent with its optical classification. The AGN indicators (broad lines, nonthermal continuum) in the upper row decrease in strength with lower X-ray luminosity and the optical spectrum becomes more and more dominated by the stars in the host galaxy (see also \cite{leh01}). However, the X-ray luminosity and hardness ratios still indicate that the X-ray flux is dominated by type-1 AGN
emission. In the lower row the X-ray luminosity still is above that of typical
starburst galaxies and in addition the X-ray spectrum gets significantly harder, with hardness ratios above 0, indicating strongly absorbed X-ray continua. These objects are classified as Seyfert-2 galaxies, although their optical spectra do not necessarily reveal any AGN features. The spectrum in the lower left shows a Seyfert-2 galaxy with significant X-ray absorption and an AGN-type luminosity. The latter spectrum is characteristic for the bulk of the detected galaxies, which show either no or very faint high excitation lines indicating the AGN nature of the object, so that we have to resort to a combination of optical and X-ray diagnostics to classify them as AGN.

\begin{figure}
  \includegraphics[height=.3\textheight]{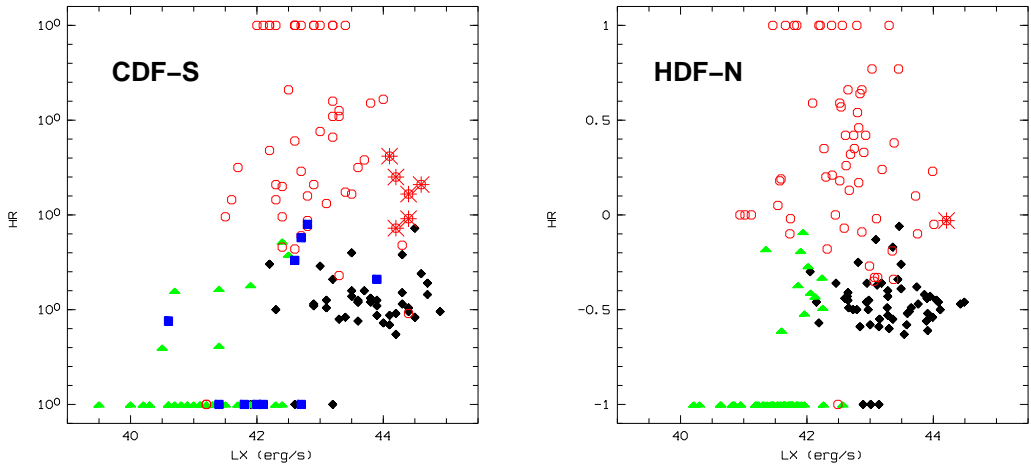}
  \caption{Hardness ratio versus rest frame luminosity in the total 0.5-10 keV band. Objects are coloured according to their X-ray/optical classification: filled black diamonds correspond to type-1 AGN, open red hexagons to type-2 AGN, green triangles to galaxies and blue squares to extended X-ray sources. The large asterisks indicates type-2 QSOs. A critical density universe with 
$H_0$ = 50 km s$^{-1}$ Mpc$^{-1}$ has been adopted. Luminosities are not corrected for possible intrinsic absorption.
 }
\end{figure}

\section{X-ray/optical classification}

Following \cite{ros02,has02} we show in Figure 3 {left} the hardness ratio as a function of the luminosity in the 0.5-10 keV band for 165 sources for which we have optical spectra and rather secure classification in the CDF-S \cite{szo03}.  The X-ray luminosities are not corrected for internal absorption and are computed in a critical density universe with $H_0$ = 50 km s$^{-1}$ Mpc$^{-1}$. Different source types are clearly segregated in this plane. Type-1 AGNs (black diamonds) have luminosities typically above $10^{42}$ erg s$^{-1}$, with hardness ratios in a narrow range around HR$\approx$-0.5. Type-2 AGN are skewed towards significantly higher hardness ratios (HR>0), with (absorbed) luminosities in the range $10^{41-44}$ erg s$^{-1}$. Direct spectral fits of the XMM-Newton and Chandra spectra clearly indicate that these harder spectra are due to neutral gas absorption and not due to a flatter intrinsic slope (see \cite{mai02,bau03}). Therefore the unabsorbed, intrinsic luminosities of type-2 AGN would fall in the same range as those of type-1's. 

In Figure 3 we also indicate the type-2 QSOs (asterisks), the first one of which was discovered in the CDF-S \cite{nor02}. In the meantime, more examples have been found in the CDF-S and elsewhere \cite{ste02}. It is interesting to note that no high-luminosity, very hard sources exist in this diagram. This is a selection effect of the pencil beam surveys: due to the small solid angle, the rare high luminosity sources are only sampled at high redshifts, where the absorption cut-off of type-2 AGN is redshifted to softer X-ray energies. Indeed, the type-2 QSOs in this sample are the objects at $L_X>10^{44}$ erg s$^{-1}$ and HR$>$-0.2. The type-1 QSO in this region of the diagram is a BAL QSO with significant intrinsic absorption.

About 10\% of the objects have optical spectra of normal galaxies (marked with triangles), luminosities below $10^{42}$ erg s$^{-1}$ and very soft X-ray spectra (several with HR=-1), as expected in the case of starbursts or thermal halos. Those at $L_X<10^{41.5}$ erg s$^{-1}$  and HR$<$-0.7 are at particularly low redshifts.  However, a separate subset has harder spectra (HR$>$-0.5), and luminosities $>10^{41}$ erg s$^{-1}$. In these galaxies the X-ray emission is likely due to a mixture of low level AGN activity and a population of low mass X-ray binaries (see \cite{bar01}).  Therefore the deep Chandra and XMM-Newton surveys detect for the first time the population of normal starburst galaxies out to intermediate redshifts \cite{mus00,gia01,leh02}. These galaxies might become an important means to study the star formation history in the universe completely independently from optical/UV, sub-mm or radio observations.

Figure 3 (right) shows the same diagram for the spectroscopic identifications in the HDF-N \cite{bar02}. While the authors give only purely optical classification information for the X-ray counterparts (basically "galaxy" or "broad line object"), we have applied the above X-ray/optical classification scheme also to their catalogue. The corresponding diagram shows basically the same features: the X-rays show that type-1 (mainly broad-line) AGN cluster around HR$\approx$-0.5 and break the degeneracy between type-2 Seyferts and normal galaxies.

\begin{figure}
  \includegraphics[height=.3\textheight]{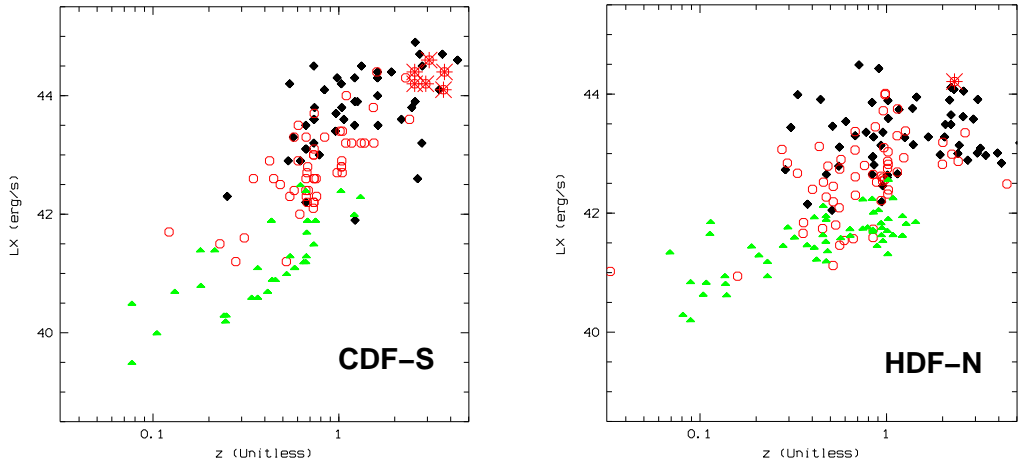}
  \caption{X-ray luminosity versus redshift magnitude for the CDF-S-sources (left) and the HDF-N sources (right). Symbols are the same as in Figure 3.}
\end{figure}

There is, however, one significant difference between the two identification samples: while the CDF-S identification catalogue \cite{szo03} includes 6 type-2 QSOs (see figure 3 (left)), which all are characterised by strong and narrow UV emission lines (Lyman-$\alpha$, CIV etc.) with almost absent continuum, the corresponding HDF-N catalogue \cite{bar02} lists only one possible type-2 QSO. A closer look to Figure 4 and the optical magnitudes of the CDF-S type-2 QSO shows that these objects are predominantly detected at redshifts z$>$2 and optical magnitudes R$>$ 24. One reason for the relative absence of this population in the HDF-N could be that a smaller number of identifications at R$>24$ exist in this survey compared to the CDF-S. However, a true field to field variation (cosmic variance) cannot be ruled out. At least in the CDF-S there is no significant variation of the ratio of type-1 to type-2 objects over the redshift range z=0.5-4.

\section{The redshift distribution}

The current spectroscopic/photometric completeness of the CDF-S and HDF-N identifications allows to compare the observed redshift distribution with predictions from X--ray background population synthesis models \cite{gil01}, which, due to the saturation of the QSO evolution predict a maximum at redshifts around z=1.5. Figure 5 shows two predictions of the redshift distribution from the Gilli et al. model for a flux limit of $2.3 \times 10^{-16}$ erg cm$^{-2}$ s$^{-1}$ in the 0.5-2 keV band with different assumptions for the high-redshift evolution of the QSO space density. The two models have been normalized at the peak of the distribution.

\begin{figure}
  \includegraphics[height=.25\textheight]{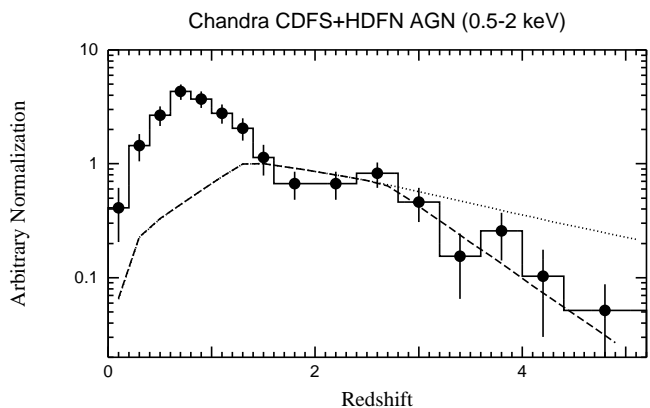}
  \caption{Redshift distribution of 243 AGN selected in the 0.5-2 keV band from the inner 10 arcmin radius of the Chandra CDF-S and HDF-N survey samples (solid circles and histogram), compared to model predictions from population synthesis models \cite{gil01}. The dashed line shows the prediction for a model, where the comoving space density of high-redshift QSO follows the decline above z=2.7 observed in optical samples \cite{ssg95,fan01}). The dotted line shows a prediction with a constant space density for $z>1.5$. The two model curves have been normalized to their peak at z=1, while the observed distribution has been normalized to roughly fit the models in the redshift range 1.5--2.5}
\end{figure}

The actually observed redshift distribution of AGN selected from the HDF-N and CDF-S Chandra deep survey samples at off-axis angles below 10 arcmin and in the 0.5-2 keV band has been arbitrarily normalized to roughly fit the population synthesis models in the redshift range 1.5 - 2.5 and shown in Fig. 5 as histogram and data points. In the redshift range below 1.5 it is radically different from the prediction, with a peak at a redshift at z$\approx$0.7. This low redshift peak is dominated by Seyfert galaxies with X-ray luminosities in the range $L_X = 10^{42-44}$ erg/s. Since the peak in the observed redshift distribution is expected at the redshift, where the strong positive evolution of AGN terminates, we can conclude that the evolution of Seyfert galaxies is significantly different from that of QSOs, with their evolution saturating around a redshift of 0.7, compared to the much earlier evolution of QSOs which saturates at z$\approx$1.5. The statistics of the two samples is now sufficient to rule out the constant space density model at redshifts above 3, clearly indicating a decline of the X-ray selected QSO population at high redshift consistent with the optical findings. However, the statistical errors and the likely spectroscopic incompleteness still preclude a more accurate determination. 

\begin{figure}
  \includegraphics[height=.3\textheight]{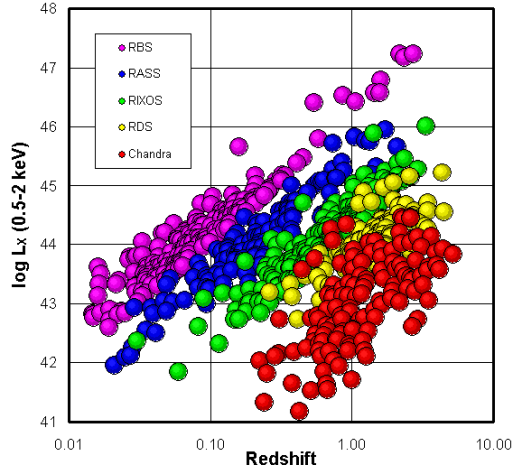}
  \caption{Luminosity as a function of redshift for different survey samples of type-1 AGN selected in the 0.5-2 keV band. The ROSAT surveys correspond to the same samples utilized in \cite{miy00}: ROSAT Bright Survey (RBS; \cite{schw00}; magenta), ROSAT All-Sky-Survey northern selected areas (RASS; \cite{app98}; blue), ROSAT International X-ray Optical Survey (RIXOS; \cite{mas00}; green), ROSAT Deep Surveys (NEP, \cite{bow96}; Marano field, \cite{zam99}; UKDS, \cite{mch98}; Lockman Hole, \cite{schm98,leh01}; yellow). The Chandra objects are from the CDF-S and HDF-N (\cite{szo03,bar02}; red).}
\end{figure}
  
\section{The ROSAT/Chandra luminosity function}

To investigate in more detail, where the difference between the Chandra objects and the predictions, which are based on the ROSAT data originate from, we have calculated preliminary luminosity functions. For the first time the Chandra deep survey data were merged with the whole body of previously identified ROSAT AGN samples, used in \cite{miy00} to compute the AGN luminosity function. To make the analysis as complete and homogeneous as possible, we have selected only the type-1 AGN in all samples and treated only the detections and X-ray fluxes in the 0.5-2 keV band. Figure 6 shows the X-ray "Hubble-Diagram" for the different samples, covering an unprecedented six orders of magnitude in flux limit and seven orders of magnitude in survey solid angle between the ROSAT Bright Survey and the Chandra Deep Survey. This diagram shows, that the new Chandra sources
are predominantly Seyfert galaxies at a median luminosity of $\approx10^{43}$ erg s$^{-1}$ and a median redshift around 0.7. 

A preliminary luminosity function was calculated using the $V/V_a$ method and is shown in two redshift shells (z=0.015-0.2 and z=1.6-2.3) in Figure 7 (left). The shape of the two luminosity function is significantly different, so that the cosmological evolution can be described neither by pure luminosity nor pure density evolution. The surprising result is, however, that the high-redshift luminosity function is almost horizontal at luminosities below $\approx10^{44}$ erg s$^{-1}$ and approaches the local space density in the Seyfert range. The strong positive density evolution, well known from previous AGN samples in the optical, radio and X-ray range, therefore only holds for relatively luminous AGN (i.e. QSOs), while the lower luminosity AGN (Seyfert galaxies) show much less or even negative density evolution. 

\begin{figure}
  \includegraphics[height=.24\textheight]{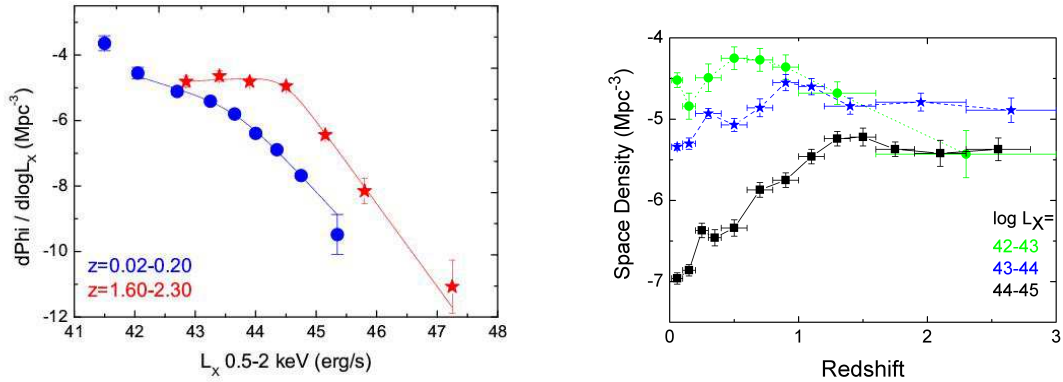}
  \caption{Left: Luminosity function of type-1 AGN selected in the 0.5-2 keV band in two redshift shells: z=0.02-0.2 \cite{miy01} and z=1.6-2.3 (this work). Right: Space density as a function of redshift for different luminosity classes.}
\end{figure}

To illustrate the different evolutionary behaviour for different luminosity classes in more detail we calculated space densities as a function of redshift for different luminosity classes: $log(L_X) =$42-43, 43-44 and 44-45, shown in Figure 7 (right). While the evolution of the highest luminosity class (44-45), the QSOs, follows very well the known strong positive evolution with an increase of almost two orders of magnitude in space density, saturating at z$\approx1.5$, the evolution of lower luminosity classes is weaker and saturates at significantly later redshifts. The highest space density is achieved for the Seyferts of luminosity class 42-43 at redshifts around 0.7 at a space density about a factor of 10 higher than that of QSOs. Beyond z=0.7 there is a significant decline of the Seyfert space density. This is the reason, why the Chandra deep surveys are dominated by this type of object in this redshift shell and not, as originally expected by Seyfert galaxies at higher redshifts.

These, still preliminary, new results paint a dramatically different evolutionary picture for low-luminosity AGN compared to the high-luminosity QSOs. While the rare, high-luminosity objects can form and feed very efficiently rather early in the universe, the bulk of the AGN has to wait much longer to
grow. This could indicate two modes of accretion and black hole growth with different accretion efficiency, as e.g. proposed in \cite{dus02}.  
The late evolution of the low-luminosity Seyfert population is very similar to 
that which is required to fit the Mid-infrared source counts and background (see e.g. \cite{fra02}), however, contrary to what has been assumed by Franceschini et al., this evolution applies to all low-luminosity AGN (type-1 and type-2). 

These results, however, have still to be taken with a grain of salt. First, the spectroscopic incompleteness in the Chandra samples is still substantial and before a formal publication of all spectroscopic and photometric redshifts is available, it is too early to draw final conclusions. Also, the derivation of the luminosity function and its evolution still needs to be confirmed by more modern methods (see e.g. \cite{schm99,miy01}).

\begin{theacknowledgments}

I thank the Chandra Deep Field South team for the good cooperation and the permission to use some data in advance of publication. I am grateful to Maarten Schmidt and Takamitsu Miyaji for the collaboration on the preliminary Chandra/ROSAT luminosity function. 
  
\end{theacknowledgments}

\bibliographystyle{aipproc}

\end{document}